\documentclass[conference,a4paper]{IEEEtran}

\IEEEoverridecommandlockouts

\usepackage[utf8]{inputenc} 
\usepackage[T1]{fontenc}
\usepackage{url}
\usepackage{ifthen}
\usepackage{cite}
\usepackage[cmex10]{amsmath} 

\usepackage[IEEEtran,notheorems,graphics]{research17}
\usepackage{booktabs} 
\usepackage{enumitem}
\usepackage{mathtools}
\usepackage{cite}
\usepackage[lined,boxed,commentsnumbered,linesnumbered,ruled]{algorithm2e}
\usepackage{tikz} 
\usetikzlibrary{shapes, patterns,decorations.text, decorations.pathreplacing}

\newtheorem{lemma}{\mylemmaname}
\newtheorem{theorem}{\mytheoremname}
\newtheorem{definition}{\mydefinitionname}

\newtheorem{corollary}{\mycorollaryname}
\newtheorem{example}{\myexamplename}

\newtheorem{claim}{\myclaimname}
\newtheorem{conjecture}{\myconjecturename}
\usepackage{multirow}
\usepackage{array}
\newcolumntype{C}[1]{>{\centering\arraybackslash}p{#1}}
\makeatletter
\renewcommand*\env@matrix[1][*\c@MaxMatrixCols c]{%
  \hskip -\arraycolsep
  \let\@ifnextchar\new@ifnextchar
  \array{#1}}
\makeatother
\usepackage[capitalize]{cleveref}
\crefname{equation}{\unskip}{\unskip}
\crefname{claim}{Claim}{Claims} 
\allowdisplaybreaks

\newcommand{\Nat}[1]{\Naturals_{#1}} 
\newcommand{\code}[1]{\mathcal{#1}} 
\renewcommand{\vect}[1]{\vectg{#1}} 
\renewcommand{\mat}[1]{\bm{#1}} 
\renewcommand{\r}{\color{red}}
\renewcommand{\b}{\color{blue}}

\begin{document}
\title{An MDS-PIR Capacity-Achieving Protocol for Distributed Storage Using Non-MDS Linear Codes}



\author{\IEEEauthorblockN{Hsuan-Yin Lin\IEEEauthorrefmark{2}, Siddhartha Kumar\IEEEauthorrefmark{2}, Eirik
    Rosnes\IEEEauthorrefmark{2}, and Alexandre Graell i Amat\IEEEauthorrefmark{3}}
  \IEEEauthorblockA{\IEEEauthorrefmark{2}Simula@UiB, N--5020 Bergen, Norway}
  \IEEEauthorblockA{\IEEEauthorrefmark{3}Department of Electrical Engineering, Chalmers University of Technology,
    SE--41296 Gothenburg, Sweden}\thanks{This work was partially funded by the Research Council of Norway (grant
    240985/F20) and the Swedish Research Council (grant \#2016-04253).}}



\maketitle
\begin{abstract}
  We propose a private information retrieval (PIR) protocol for distributed storage systems with noncolluding nodes
  where data is stored using an arbitrary linear code. An expression for the PIR rate, i.e., the ratio of the amount of
  retrieved data per unit of downloaded data, is derived, and a necessary and a sufficient condition for codes to
  achieve the maximum distance separable (MDS) PIR capacity are given. The necessary condition is based on the
  generalized Hamming weights of the storage code, while the sufficient condition is based on code automorphisms.  We
  show that cyclic codes and Reed-Muller codes satisfy the sufficient condition and are thus MDS-PIR capacity-achieving.
  %
\end{abstract}



\section{Introduction}
\label{sec:introduction}

Private information retrieval (PIR) was first addressed by Chor \emph{et al.} in the computer science community
\cite{ChorKushilevitzGoldreichSudan95_1}. A PIR protocol allows the users to privately retrieve any requested file
stored in a set of servers (referred to as nodes in the sequel) without revealing to the nodes which file is actually
being downloaded. The efficiency of a PIR protocol is measured in terms of the overall communication cost, defined as
the sum of the upload and the download cost, and the goal is to design a protocol that minimizes it. Recently, PIR for
coded distributed storage systems (DSSs) where data is encoded by a linear code and then stored across nodes has
attracted a great deal of attention
\cite{IshaiKushilevitzOstrovskySahai04_1,ShahRashmiRamchandran14_1,ChanHoYamamoto15_1,TajeddineElRouayheb16_1,
  KumarRosnesGraell17_1,FreijHollantiGnilkeHollantiKarpuk17_1}.

Assuming that the size of the files stored in the DSS is much larger than the number of files stored, the upload cost is
small compared to the download cost \cite{ChanHoYamamoto15_1,TajeddineElRouayheb16_1}, and thus it can be ignored.
The PIR rate is then defined as the amount of information retrieved per downloaded symbol and is the measure of
efficiency used in the information theory community. For the so-called uncoded PIR problem, where the system can be seen
as a coded DSS using a repetition code, the authors in \cite{SunJafar17_1,SunJafar18_2} derived an exact expression for
the maximum possible PIR rate over all possible PIR protocols, i.e., the \emph{PIR capacity}.  A closed-form expression
for the coded PIR capacity when no nodes collude was derived in \cite{BanawanUlukus18_1} for the case where data is
encoded by a maximum distance separable (MDS) code and then stored. The PIR capacity in this case is usually referred to
as \emph{MDS-PIR} capacity. Most of the earlier works focus on studying PIR schemes for DSSs where data is stored
using an MDS code.
A PIR protocol for DSSs where data is stored using an arbitrary linear code was considered in
\cite{KumarRosnesGraell17_1} for the case of noncolluding nodes, and it was shown that the asymptotic MDS-PIR capacity
(assuming that an infinite number of files are stored) can be achieved even when the underlying code is non-MDS. PIR
with linear codes for the case of colluding nodes was addressed in
\cite{FreijHollantiGnilkeHollantiKarpuk17_1,KumarLinRosnesGraell17_1sub,FreijHollantiGnilkeHollanti17_1sub}.  A
conjecture for the MDS-PIR capacity in the colluding case was stated in \cite{FreijHollantiGnilkeHollantiKarpuk17_1},
but disproved for $2$ files in \cite{SunJafar17_2}. However, in other cases (e.g., in the asymptotic case), it is still
open.

In this paper, we propose a PIR protocol for DSSs where data is stored using an arbitrary linear code for the case of
noncolluding nodes. Furthermore, we investigate which classes of codes can achieve the MDS-PIR capacity with the
proposed protocol. Specifically, we derive an expression for the PIR rate  by exploiting a number of
coordinate sets containing information sets of the underlying storage code, and define a class of MDS-PIR
capacity-achieving codes, which includes MDS codes.
We also provide a necessary  and a sufficient condition for a code to achieve the MDS-PIR capacity with the given PIR protocol. The
necessary condition is connected to the generalized Hamming weights of the storage code, while the sufficient condition
is related to code automorphisms.
We show that cyclic codes and Reed-Muller (RM) codes satisfy the sufficient condition and are thus MDS-PIR
capacity-achieving codes. In the following, all proofs are omitted due to lack of space. The proofs can be found in the
extended version \cite{KumarLinRosnesGraell17_1sub}.

Notation: We use $\Nat{}$ for the set of all positive integers, $\Nat{a}\eqdef\{1,2,\ldots,a\}$, and
$\Nat{n_1:n_2}\eqdef\{n_1,n_1+1,\ldots,n_2\}$ for two positive integers $n_1\leq n_2$, $n_1,n_2\in\Nat{}$.
Vectors are denoted by lower case bold letters, matrices by upper case bold letters, and sets by calligraphic upper case
letters, e.g., $\bm x$, $\bm X$, and $\mathcal{X}$ denote a vector, a matrix, and a set, respectively. $\code{C}$ will
denote a linear code over the finite field $\GF(q)$. We use the customary code parameters $[n,k]$ or
$[n,k,d^\code{C}_\mathsf{min}]$ to denote a code $\code{C}$ of blocklength $n$, dimension $k$, and minimum Hamming
distance $d^\code{C}_\mathsf{min}$.
$\trans{(\cdot)}$ represents the transpose of its argument, while $(\bm X_1|\ldots|\bm X_a)$ represents the horizontal
concatenation of the matrices $\bm X_1,\ldots,\bm X_a$, all with the same number of rows.  The function
$\mathsf{H}(\cdot)$ represents the entropy of its argument and $\chi(\vect{x})$ denotes the support of a vector
$\vect{x}$. Subscripts may be omitted if the arguments we refer to are contextually unambiguous.

\section{Preliminaries and System Model}
\label{Sec:DefandPrelim}

In this section, we first review some notions in coding theory and then give the system model and the privacy model.

\subsection{Definitions}

\begin{definition}
  \label{def:infoS}
  Let $\code{C}$ be an $[n,k]$ code with generator matrix $\mat{G}^\code{C}$, and denote by
  $\mat{G}^\code{C}|_{\set{I}}$ the matrix consisting of the columns of $\mat{G}^\code{C}$ indexed by $\set{I}$. A set
  of coordinates of $\code{C}$, $\set{I}\subseteq\Nat{n}$, of size $k$ is said to be an \emph{information set} if and
  only if $\mat{G}^\code{C}|_\set{I}$ is invertible.
\end{definition}

\begin{definition}
  \label{def:suppD}
  Let $\code{D}$ be a subcode of an $[n,k]$ code $\code{C}$. The \emph{support} of $\code{D}$ is defined as
  \begin{equation*}
    \chi(\code{D})\eqdef\{j\in\Nat{n}\colon\exists\,\vect{x}=(x_1,\ldots,x_n)\in\code{D}, x_j\neq 0\}.
  \end{equation*}
\end{definition}

\begin{definition}[{Generalized Hamming weight \cite{wei91_1}}]
  \label{def:sth-GHW}
  The $s$-th generalized Hamming weight of an $[n,k]$ code $\code{C}$, denoted by $d_s^{\code{C}}$, $s\in\Nat{k}$, is
  defined as the cardinality of the smallest support of an $s$-dimensional subcode of $\code{C}$, i.e.,
  \begin{equation*}
    d_s^{\code{C}}\eqdef\min\bigl\{\card{\chi(\code{D})}\colon\code{D}\text{ is an } [n,s] \text{ subcode of }
    \code{C}\bigr\}.
  \end{equation*}
\end{definition}

\subsection{System Model}
\label{sec:SystemModel}

We consider a DSS that stores $f$ files $\mat{X}^{(1)},\ldots,\mat{X}^{(f)}$, where each file
$\mat{X}^{(m)}=(x_{i,j}^{(m)})$, $m\in\Nat{f}$, can be seen as a $\beta\times k$ matrix over $\GF(q)$ with
$\beta,k \in\Nat{}$. Each file 
is encoded using a linear code as follows. Let $\vect{x}^{(m)}_i=\bigl(x^{(m)}_{i,1},\ldots,x^{(m)}_{i,k}\bigr)$,
$i\in\Nat{\beta}$, be a message vector corresponding to the $i$-th row of $\mat{X}^{(m)}$. Each $\vect{x}^{(m)}_i$ is
encoded by an $[n,k]$ code $\code{C}$ over $\GF(q)$ into a length-$n$ codeword
$\vect{c}^{(m)}_i=\bigl(c^{(m)}_{i,1},\ldots,c^{(m)}_{i,n}\bigr)$, where $c_{i,j}^{(m)}\in\GF(q)$, $j \in\Nat{n}$. The
$\beta f$ generated codewords $\vect{c}_i^{(m)}$ are then arranged in the array
$\mat{C}=\trans{\bigl(\trans{(\mat{C}^{(1)})}|\ldots|\trans{(\mat{C}^{(f)})}\bigr)}$ of dimensions $\beta f \times n$,
where $\mat{C}^{(m)}=\trans{\bigl(\trans{(\vect{c}^{(m)}_1)}|\ldots|\trans{(\vect{c}^{(m)}_{\beta})}\bigr)}$ for
$m\in\Nat{f}$. For a given column $j$ of $\mat{C}$, we denote the column vector
$\trans{\bigl(c_{1,j}^{(m)},\ldots,c_{\beta,j}^{(m)}\bigr)}$ as a coded chunk pertaining to file $\mat{X}^{(m)}$. The
$f$ coded chunks in column $j$ are stored on the $j$-th storage node, $j\in\Nat{n}$.

\subsection{Privacy Model}
\label{sec:privacy}

We consider a DSS where a node may act as spy. It is assumed that the remaining nonspy nodes do not collaborate with the
spy node. The scenario of one spy node is analogous to having a system with no colluding nodes. To retrieve file
$\mat{X}^{(m)}$ from the DSS, the user sends a $d \times \beta f$ query matrix $\mat{Q}^{(l)}$ over $\GF(q)$, for some
integer $d$, to the $l$-th node for all $l\in\Nat{n}$. In response to the received query, node $l$ sends the response
vector $\vect{r}_l$, which is a deterministic function of $\mat{Q}^{(l)}$ and the code symbols stored in the node $l$,
back to the user.
\begin{definition}
  \label{Def:cond}
  Consider a DSS with $n$ nodes storing $f$ files in which a single node acts as spy. A user who wishes to retrieve the
  $m$-th file sends the queries $\mat{Q}^{(l)}$, $l\in\Nat{n}$, to the storage nodes, which return the responses
  $\vect{r}_l$. This scheme achieves perfect information-theoretic PIR if and only if
  \begin{subequations}
  \begin{IEEEeqnarray}{rCl}
    \text{Privacy:}\qquad &&\mathsf{H}\bigl(m|\mat{Q}^{(l)}\bigr)=\mathsf{H}(m),\,\forall\,l\in\Nat{n};\label{eq:cond1}
    \\
    \text{Recovery:}\qquad&&\mathsf{H}\bigl(\mat{X}^{(m)}|\vect{r}_1,\ldots,\vect{r}_n\bigr)=0.\label{eq:cond2}
  \end{IEEEeqnarray}
   \end{subequations}
\end{definition}

Queries satisfying \eqref{eq:cond1} ensure that the spy node is not able to determine which file is being downloaded by
the user. The recovery constraint in \eqref{eq:cond2} ensures that the user is able to recover the requested file from
the responses sent by the DSS.

\begin{definition}
  \label{def:def_PIRrate}
  The PIR rate of a PIR protocol, denoted by $\const{R}$, is the amount of information retrieved per downloaded symbol,
  i.e., $\const{R}\eqdef\frac{\beta k}{\const{D}}$, where $\const{D}$ is the total number of downloaded symbols for the  
  retrieval of a single file.
\end{definition}

We will write $\const{R}(\mathcal C)$ to highlight that the PIR rate depends on the underlying storage code
$\code{C}$. The maximum achievable PIR rate 
is the PIR capacity. It was shown in \cite{BanawanUlukus18_1} that for the
noncolluding case and for a given number of files $f$ stored using an $[n,k]$ MDS code, the MDS-PIR capacity, denoted by
$\const{C}_f$, is
\begin{equation}
  \const{C}_f\eqdef\frac{n-k}{n}\inv{\left[1-\Bigl(\frac{k}{n}\Bigr)^f\right]}.
  \label{eq:PIRcapacity}  
\end{equation}

\section{Capacity-Achieving PIR Protocol}
\label{sec:file-dep-PIR}

In this section, we propose a PIR protocol that achieves the MDS-PIR capacity for the scenario of
noncolluding nodes. The protocol is inspired by the protocol introduced in \cite{BanawanUlukus18_1}.

\subsection{PIR Achievable Rate Matrix}
\label{sec:PIRachievable-rate-matrix}

In \cite{SunJafar17_1}, the concept of exploiting \emph{side information} for PIR problems was introduced. 
By side information we mean additional redundant symbols not related to the requested file but downloaded by the user in
order to maintain privacy. These symbols can be exploited by the user to retrieve the requested file from the responses
of the storage nodes. In \cite[Sec.~V.A]{BanawanUlukus18_1}, it was shown that for a $[5,3,3]$ MDS storage code, the
side information is decoded by utilizing other code coordinates forming an information set in the code array. For
instance, the authors chose the $\nu=5$ information sets $\set{I}_1=\{1,2,3\}$, $\set{I}_2=\{1,4,5\}$,
$\set{I}_3=\{2,3,4\}$, $\set{I}_4=\{1,2,5\}$, and $\set{I}_5=\{3,4,5\}$ of the $[5,3,3]$ MDS code in their PIR
achievable scheme. Observe that in $\{\set{I}_i\}_{i\in\Nat{5}}$ each coordinate of the $[5,3,3]$ code appears exactly
$\kappa=3$ times. This motivates the following definition.

\begin{definition}
  \label{def:PIRachievable-rate-matrix}
  Let $\code{C}$ be an arbitrary $[n,k]$ code. A $\nu\times n$ binary matrix $\mat{\Lambda}_{\kappa,\nu}(\code{C})$ is
  said to be a \emph{PIR achievable rate matrix} for $\code{C}$ if the following conditions are satisfied.
  \begin{enumerate}
  \item \label{item:1} The Hamming weight of each column of $\mat{\Lambda}_{\kappa,\nu}$ is $\kappa$, and
  \item \label{item:2} for each matrix row $\vect{\lambda}_i$, $i\in\Nat{\nu}$, $\chi(\vect{\lambda}_i)$ always contains
    an information set.
  \end{enumerate}
  In other words, each coordinate $j$ of $\code{C}$, $j\in\Nat{n}$, appears exactly $\kappa$ times in
  $\{\chi(\vect{\lambda}_i)\}_{i\in\Nat{\nu}}$, and every set $\chi(\vect{\lambda}_i)$ contains an information set.
\end{definition}

\begin{lemma}
  \label{lem:PIRrate_upper-bound}
  If a matrix $\mat{\Lambda}_{\kappa,\nu}(\code{C})$ exists for an $[n,k]$ code $\code{C}$, then we have
  \begin{equation*}
    \frac{\kappa}{\nu}\geq \frac{k}{n},
  \end{equation*}
  where equality holds if $\chi(\vect{\lambda}_i)$, $i\in\Nat{\nu}$, are all information sets.
\end{lemma}

\begin{example}
  \label{ex:n5k3_bad}
  Consider the $[5,3,2]$ code with generator matrix
  \begin{equation*}
    \mat{G}=
    \begin{pmatrix}
      1 & 0 & 0 & 1 & 0
      \\
      0 & 1 & 0 & 1 & 0
      \\
      0 & 0 & 1 & 0 & 1
    \end{pmatrix}.
  \end{equation*}
  One can easily verify that
  \begin{equation*}
    \mat{\Lambda}_{2,3}=
    \begin{pmatrix}
      0 & 1 & 1 & 1 & 1
      \\
      1 & 0 & 0 & 1 & 1
      \\
      1 & 1 & 1 & 0 & 0
    \end{pmatrix}
  \end{equation*}
  is a PIR achievable rate matrix for this code.
\end{example}

Before we state our main results, in order to clearly illustrate the PIR protocol, we first introduce the
following definition.
\begin{definition} \label{def:InterferenceAB}
  \label{def:PIRinterference-matrices}
  For a given $\nu\times n$ PIR achievable rate matrix $\mat{\Lambda}_{\kappa,\nu}(\code{C})=(\lambda_{u,j})$, we define
  the PIR interference matrices $\mat{A}_{\kappa{\times}n}=(a_{i,j})$ and $\mat{B}_{(\nu-\kappa){\times}n}=(b_{i,j})$
  with
  \begin{IEEEeqnarray*}{rCl}
    a_{i,j}& \eqdef &u \text{ if } \lambda_{u,j}=1,\,\forall\,j\in\Nat{n},\,i\in\Nat{\kappa},u\in\Nat{\nu},
    \\
    b_{i,j}& \eqdef &u \text{ if } \lambda_{u,j}=0,\,\forall\,j\in\Nat{n},\,i\in\Nat{\nu-\kappa},u\in\Nat{\nu}.
  \end{IEEEeqnarray*}
\end{definition}

Note that in \cref{def:InterferenceAB}, for each $j\in\Nat{n}$, distinct values of $u\in\Nat{\nu}$ should be assigned
for all $i$. Thus, the assignment is not unique in the sense that the order of the entries of each column of $\mat{A}$
and $\mat{B}$ can be permuted. For $j\in\Nat{n}$, let $\set{A}_j\eqdef\{a_{i,j}\colon i\in\Nat{\kappa}\}$ and
$\set{B}_j\eqdef\{b_{i,j}\colon i\in\Nat{\nu-\kappa}\}$. Note that the $j$-th column of $\mat{A}$ contains the row
indices of $\mat{\Lambda}$ whose entries in the $j$-th column are equal to $1$, while $\mat{B}$ contains the remaining
row indices of $\mat{\Lambda}$. Hence, it can be observed that $\set{B}_j=\Nat{\nu}\setminus\set{A}_j$,
$\forall\,j\in\Nat{n}$.
\begin{definition}
  \label{def:aSet_A}
  By $\set{S}(a|\mat{A}_{\kappa \times n})$ we denote the set of column coordinates of matrix
  $\mat{A}_{\kappa{\times}n}=(a_{i,j})$ for which at least one of its entries is equal to $a$, i.e.,
  \begin{equation*}
    \set{S}(a|\mat{A}_{\kappa{\times}n})\eqdef\{j\in\Nat{n}\colon\exists\,a_{i,j}=a,i\in\Nat{\kappa}\}.
  \end{equation*}
\end{definition}

The following claim can be directly verified.
\begin{claim}
  \label{clm:property_PIRinterference-matrices}
  $\set{S}(a|\mat{A}_{\kappa\times n})$ contains an information set $\forall\,a\in\Nat{\nu}$. Moreover, for an arbitrary
  entry $b_{i,j}$ of $\mat{B}_{(\nu-\kappa)\times n}$,
  $\set{S}(b_{i,j}|\mat{A}_{\kappa\times n})\subseteq\Nat{n}\setminus\{j\}$ and it must contain an information set.
\end{claim}

We illustrate the previous points in the following example.
\begin{example}
  \label{ex:A-B_n5k3_bad}
  Continuing with \cref{ex:n5k3_bad} and following \cref{def:PIRinterference-matrices}, we obtain
  \begin{IEEEeqnarray*}{rCl}
    \mat{A}_{2\times 5}=
    \begin{pmatrix}
      2 &1 &1 &1 & 1
      \\
      3 &3 &3 &2 & 2
    \end{pmatrix}
    \text{ and } \mat{B}_{1\times 5}=
    \begin{pmatrix}
      1 &2 &2 &3 & 3
    \end{pmatrix}
  \end{IEEEeqnarray*}
  for $\mat{\Lambda}_{2,3}$. One can see that $\set{A}_j\cup\set{B}_j=\Nat{3}$, $\forall\,j\in\Nat{5}$. Moreover, for
  instance, take $a=1$, then $\set{S}(1|\mat{A}_{2\times 5})=\{2,3,4,5\}$ contains an information set of the $[5,3,2]$
  systematic code of Example~\ref{ex:n5k3_bad}.
\end{example}

Now consider the two matrices
\begin{IEEEeqnarray*}{rCl}
  &\begin{pmatrix}
    c^{(m)}_{\mu + {\r a_{1,1}},1} & c^{(m)}_{\mu+{\r a_{1,2}},2} &\cdots & c^{(m)}_{\mu+{\r a_{1,n}},n}
    \\
    \vdots
    \\
    c^{(m)}_{\mu+{\r a_{\kappa,1}},1} & c^{(m)}_{\mu+{\r a_{\kappa,2}},2} &\cdots & c^{(m)}_{\mu+{\r a_{\kappa,n}},n}
  \end{pmatrix}&
  \text{and}\\[2mm]
  &\begin{pmatrix}
    c^{(m)}_{\mu+{\b b_{1,1}},1} & c^{(m)}_{\mu+{\b b_{1,2}},2} &\cdots & c^{(m)}_{\mu+{\b b_{1,n}},n}
    \\
    \vdots
    \\
    c^{(m)}_{\mu+{\b b_{\nu-\kappa,1}},1} & c^{(m)}_{\mu+{\b b_{\nu-\kappa,2}},2} &\cdots
    & c^{(m)}_{\mu+{\b b_{\nu-\kappa,n}},n}
  \end{pmatrix}&
\end{IEEEeqnarray*}
of code symbols of the $m$-th file, where $\mu \in \Nat{\beta-\nu} \cup \{ 0 \}$. Observe that if the user knows the first matrix of code symbols, from
\cref{clm:property_PIRinterference-matrices}, since the coordinate set
$\set{S}(b_{i,j}|\mat{A}_{\kappa\times n})\subseteq\Nat{n}\setminus\{j\}$ contains an information set and the user knows
the structure of the storage code $\code{C}$, the code symbols $c^{(m)}_{\mu+b_{i,j},j}$ of the second matrix can be
obtained. Here, the entries of $\mat{A}$ and $\mat{B}$ are respectively marked in red and blue. 
The actual PIR protocol is stated below.

\subsection{PIR Protocol}
\label{sec:file-dep-PIRachievable-scheme}

The proposed PIR protocol generalizes the MDS-coded PIR protocol in \cite{BanawanUlukus18_1} to DSSs where files are
stored using an arbitrary linear code. Inspired by \cite{SunJafar17_1} and \cite{BanawanUlukus18_1}, a PIR
capacity-achievable scheme should follow two important principles: 1) enforcing file symmetry within each storage node,
and 2) exploiting side information of undesired symbols to retrieve new desired symbols.\footnote{In \cite{SunJafar17_1}
  and \cite{BanawanUlukus18_1}, a third principle was introduced, namely enforcing symmetry across storage
  nodes. However, this is in general not a necessary requirement for a feasible PIR protocol.}

The PIR achievable rate matrix $\mat{\Lambda}_{\kappa,\nu}$ for the given storage code $\code{C}$ plays a central 
role in the proposed PIR protocol. Moreover, the protocol requires $\beta=\nu^{f}$ stripes and exploits the
corresponding PIR interference matrices $\mat{A}_{\kappa\times n}$ and $\mat{B}_{(\nu-\kappa)\times n}$. Here, we simply
outline the steps of the protocol. Its detailed exposition and corresponding proofs are given in \cite[Sec.~IV-B and
App.~B]{KumarLinRosnesGraell17_1sub}. A particular example is also given in the extended version
\cite[Sec.~IV-D]{KumarLinRosnesGraell17_1sub}. Without loss of generality, we assume that the user wants to download the
first file, i.e., $m=1$. The algorithm is composed of four steps as described below.

\subsubsection*{$\mathsf{Step~1}$. Index Preparation}

For all files, the user interleaves the indices of the rows of $\mat{C}^{(m)}$ randomly and independently of each
other and generates the interleaved code array
$\mat{Y}^{(m)}=\trans{\bigl(\trans{(\vect{y}^{(m)}_1)}|\ldots|\trans{(\vect{y}^{(m)}_\beta)}\bigr)}$,
$\forall\,m\in\Nat{f}$, with rows
\begin{equation*}
  \vect{y}^{(m)}_i=\vect{c}^{(m)}_{\pi(i)}, \quad i\in\Nat{\beta},
\end{equation*}
where $\pi(\cdot):\Nat{\beta}\to\Nat{\beta}$ is a random permutation, which is privately known to the user
only. Therefore, when the user requests code symbols from each storage node, this procedure is designed to make the
requested row indices to be random and independent of the requested file index.

\subsubsection*{$\mathsf{Step~2}$. Download Symbols in the $i$-th Repetition}
\label{sec:download-symbols_ith}

The user downloads the needed symbols in $\kappa$ repetitions. In the $i$-th repetition, $i\in\Nat{\kappa}$, the user
downloads the required symbols in a total of $f$ rounds. Using the terminology in \cite{BanawanUlukus18_1}, the user
downloads two types of symbols, \emph{desired symbols}, which are directly related to the requested file index
$m=1$, and \emph{undesired symbols}, which are not related to the requested file index $m=1$, but are exploited to
decode the requested file from the desired symbols. For the desired symbols, we will distinguish between round $\ell=1$
and round $\ell\in\Nat{2:f}$.

\begin{description}[leftmargin=0cm]
\item[Undesired symbols.] The undesired symbols refer to sums of code symbols which do not contain symbols from the
  requested file. For every round $\ell$, $\ell\in\Nat{f-1}$, the user downloads the code symbols
  \begin{IEEEeqnarray*}{rCl}
    \Biggl\{
    &&\quad\sum_{m'\in\set{M}}
    y^{(m')}_{((i-1)\const{U}(f-1)+\const{U}(\ell-1))\cdot\nu+{\r a_{1,j}},j},
    \nonumber\\
    &&\ldots,\sum_{m'\in\set{M}}
    y^{(m')}_{((i-1)\const{U}(f-1)+\const{U}(\ell-1))\cdot\nu+{\r a_{\kappa,j}},j},
    \nonumber\\
    &&\ldots,\sum_{m'\in\set{M}}
    y^{(m')}_{((i-1)\const{U}(f-1)+\const{U}(\ell)-1)\cdot\nu+{\r a_{1,j}},j},
    \nonumber\\
    &&\ldots,\sum_{m'\in\set{M}}
    y^{(m')}_{((i-1)\const{U}(f-1)+\const{U}(\ell)-1)\cdot\nu+{\r a_{\kappa,j}},j}\Biggr\}
  \end{IEEEeqnarray*}
  for all $j \in \Nat{n}$ and for all possible subsets $\set{M}\subseteq\Nat{2:f}$, where $\card{\set{M}}=\ell$ and 
  $\const{U}(\ell)\eqdef\sum_{h=1}^\ell\kappa^{f-(h+1)}(\nu-\kappa)^{h-1}$.

  In contrast to undesired symbols, desired symbols are sums of code symbols which contain symbols of the requested
  file. The main idea of the protocol is that the user downloads desired symbols that are linear sums of requested
  symbols and undesired symbols from the previous round.
  
\item[Desired symbols in the first round.] In the first round, the user downloads
  $\kappa\cdot\const{U}(1)=\kappa\kappa^{f-(1+1)}(\nu-\kappa)^{1-1}=\kappa^{f-1}$ undesired symbols from each storage
  node. However, these symbols cannot be exploited directly. Hence, due to symmetry, in round $\ell=1$, the user
  downloads the $\kappa^{f-1}$ desired symbols
  \begin{equation*}
    \Bigl\{y^{(1)}_{\kappa^{f-1}({\r a_{i,j}}-1)+1,j},\ldots,y^{(1)}_{\kappa^{f-1}({\r
        a_{i,j}}-1)+\kappa^{f-1},j}\Bigr\}
  \end{equation*}
  from the $j$-th storage node, $j\in\Nat{n}$, i.e., the user also downloads $\kappa^{f-1}$ symbols for $m=1$ from each
  storage node.

\item[Desired symbols in higher rounds.] In the $(\ell+1)$-th round, $\ell\in\Nat{f-1}$, in order to exploit the side
  information, i.e., the undesired symbols from the previous round, the user downloads the symbols
  \begin{IEEEeqnarray*}{rCl}
    \Biggl\{
    &&\quad y^{(1)}_{\const{D}(\ell-1)\cdot\nu+{\r a_{i,j}},j}\nonumber\\
    &&\quad \>+
    \sum_{m'\in\set{M}_1}
    y^{(m')}_{((i-1)\const{U}(f-1)+\const{U}(\ell-1))\cdot\nu+{\b b_{1,j}},j},
    \nonumber\\[1mm]
    &&\ldots,y^{(1)}_{(\const{D}(\ell-1)+(\nu-\kappa)-1)\cdot\nu+{\r a_{i,j}},j}\nonumber\\
    &&\quad \>+
    \sum_{m'\in\set{M}_1}
    y^{(m')}_{((i-1)\const{U}(f-1)+\const{U}(\ell-1))\cdot\nu+{\b b_{\nu-\kappa,j}},j},
    \nonumber\\[1mm]    
    &&\ldots,y^{(1)}_{\bigl[\const{D}(\ell-1)+(\const{U}(\ell)-\const{U}(\ell-1))(\nu-\kappa)-1\bigr]\cdot\nu+{\r a_{i,j}},j}
    \nonumber\\[1mm]
    &&\quad \>+
    \sum_{m'\in\set{M}_1}
    y^{(m')}_{((i-1)\const{U}(f-1)+\const{U}(\ell)-1)\cdot\nu+{\b b_{\nu-\kappa,j}},j},
    \nonumber\\[1mm]
    &&\ldots,y^{(1)}_{(\const{D}(\ell)-1)\cdot\nu+{\r a_{i,j}},j}\nonumber\\
    &&\quad \>+
    \sum_{m'\in\set{M}_{\const{N}(\ell)}}
    y^{(m')}_{((i-1)\const{U}(f-1)+\const{U}(\ell)-1)\cdot\nu+{\b b_{\nu-\kappa,j}},j}\Biggr\}
  \end{IEEEeqnarray*}
  for all distinct $\ell$-sized subsets $\set{M}_1,\ldots,\set{M}_{\const{N}(\ell)}\subseteq\Nat{2:f}$, where $j \in \Nat{n}$, 
  $\const{N}(\ell)\eqdef\binom{f-1}{\ell}$, and 
  \begin{equation*}
    \const{D}(\ell)\eqdef\kappa^{f-1}+\sum_{h=1}^\ell\binom{f-1}{h}\kappa^{f-(h+1)}(\nu-\kappa)^h. 
  \end{equation*}

  This indicates that for each combination of files $\set{M}_l$, $l\in\Nat{\const{N}(\ell)}$, the user downloads
  $\bigl[\const{U}(\ell)-1-\const{U}(\ell-1)+1\bigr](\nu-\kappa)$ new desired symbols from each storage node, and since there are in total
  $\const{N}(\ell)$ combinations of files, in each round $\const{D}(\ell)-1-\const{D}(\ell-1)+1$ extra
  desired symbols are downloaded from each storage node.

\item[Exploiting the side information.] Using the fact that for a linear code $\code{C}$ any linear combination of
  codewords is also a codeword, and together with \cref{clm:property_PIRinterference-matrices}, it is not too hard to
  see that by fixing an arbitrary coordinate $j\in\Nat{n}$, there always exist some coordinates
  $\set{S}\subset\Nat{n}\setminus\{j\}$ (see \cref{clm:property_PIRinterference-matrices}) such that for a subset $\set{M}\subseteq\Nat{2:f}$ with $\card{\set{M}}=\ell$,
  the so-called \emph{aligned sum}
    \begin{IEEEeqnarray*}{rCl}
    \Biggl\{&&\sum_{m'\in\set{M}} y^{(m')}_{((i-1)\const{U}(f-1)+\const{U}(\ell-1))\cdot\nu+{\b
        b_{1,j}},j},\nonumber\\
    &&\quad \>\ldots, \sum_{m'\in\set{M}} y^{(m')}_{((i-1)\const{U}(f-1)+\const{U}(\ell)-1)\cdot\nu+{\b
        b_{\nu-\kappa,j}},j} \Biggr\}
  \end{IEEEeqnarray*}
  for $\ell\in\Nat{f-1}$ and $i\in\Nat{\kappa}$, can be decoded. Consequently, in the $(\ell+1)$-th round, from each
  storage node $j$ we can collect code symbols related to $m=1$ from the desired symbols, i.e.,
  \begin{equation*}
    \Bigl\{y^{(1)}_{\const{D}(\ell-1)\cdot\nu+{\r a_{i,j}},j},\ldots, y^{(1)}_{(\const{D}(\ell)-1)\cdot\nu+{\r
        a_{i,j}},j}\Bigr\}
  \end{equation*}
  is obtained.
\end{description}

\subsubsection*{$\mathsf{Step~3}$. Complete $\kappa$ Repetitions} The user repeats Step~2 until $i=\kappa$. We can show
that by our designed parameters $\const{U}(\ell)$ and $\const{D}(\ell)$, the user indeed downloads in total
$\beta=\nu^f$ stripes for the requested file (see \cite[App.~B]{KumarLinRosnesGraell17_1sub}).

\subsubsection*{$\mathsf{Step~4}$. Shuffling the Order of Queries to Each Node} The order of the queries to each
storage node is uniformly shuffled to prevent the storage node to be able to identify which file is requested from the
index of the first downloaded symbol.

\subsection{Achievable PIR Rate}
\label{sec:PIRrate_codes}

The PIR rate, $\const{R}(\code{C})$, of the protocol proposed in \cref{sec:file-dep-PIRachievable-scheme} for a DSS
where $f$ files are stored using an arbitrary $[n,k]$ code $\code{C}$ is given in the following theorem.
\begin{theorem}
  \label{thm:PIRachievable-rate_code}
  Consider a DSS that uses an $[n,k]$ code $\code{C}$ to store $f$ files. If a PIR achievable rate matrix
  $\mat{\Lambda}_{\kappa,\nu}(\code{C})$ exists, then
  the PIR rate 
  \begin{equation}
    \const{R}(\code{C})=\frac{(\nu-\kappa)k}{\kappa n} \inv{\left[1-\Bigl(\frac{\kappa}{\nu}\Bigr)^f\right]}
    \label{eq:PIRachievable-rate_code}    
  \end{equation}
  is achievable.
\end{theorem}
\begin{IEEEproof}
  See \cite[App.~B]{KumarLinRosnesGraell17_1sub}.
\end{IEEEproof}

We remark that from \cref{lem:PIRrate_upper-bound}, \eqref{eq:PIRachievable-rate_code} is smaller than or equal to the
MDS-PIR capacity \eqref{eq:PIRcapacity} since
\begin{IEEEeqnarray}{rCl}
  \const{R}(\code{C})& = & \frac{\frac{\nu k}{\kappa n}\Bigl[1-\frac{\kappa}{\nu}\bigr]}
  {\Bigl[1-\bigl(\frac{\kappa}{\nu}\bigr)^f\Bigr]} =\frac{\nu k}{\kappa n}\inv{\left[1+\frac{\kappa}{\nu}
      +\cdots+\Bigl(\frac{\kappa}{\nu}\Bigr)^{f-1}\right]}
  \nonumber\\
  & \leq &\inv{\left[1+\frac{k}{n} +\cdots+\Bigl(\frac{k}{n}\Bigr)^{f-1}\right]},\label{eq:upperbound_PIRrate}
\end{IEEEeqnarray}
and it becomes the MDS-PIR capacity \eqref{eq:PIRcapacity} if there exists a matrix $\mat{\Lambda}_{\kappa,\nu}$ for
$\code{C}$ with $\frac{\kappa}{\nu}=\frac{k}{n}$. The inequality in \eqref{eq:upperbound_PIRrate} follows from
\cref{lem:PIRrate_upper-bound}.

\begin{corollary}
  \label{cor:MDSPIRcapacity-achieving-matrix}
  If a  PIR achievable rate matrix $\mat{\Lambda}_{\kappa,\nu}(\code{C})$ with $\frac{\kappa}{\nu}=\frac{k}{n}$ exists 
  for an $[n,k]$ code $\code{C}$, then the MDS-PIR capacity \eqref{eq:PIRcapacity} is achievable. 
\end{corollary}

\begin{definition}
  \label{def:PIRcapacity-achieving-codes}
  A PIR achievable rate matrix $\mat{\Lambda}_{\kappa,\nu}(\code{C})$ with $\frac{\kappa}{\nu}=\frac{k}{n}$ for an $[n,k]$ code $\code{C}$ is called an 
  \emph{MDS-PIR capacity-achieving} matrix, and $\code{C}$ is referred to as an \emph{MDS-PIR capacity-achieving} code.
\end{definition}

Note that the largest achievable PIR rate in the noncolluding case where data is stored using an arbitrary linear code
is still unknown. Interestingly, it is observed from \cref{lem:PIRrate_upper-bound} and \eqref{eq:upperbound_PIRrate}
that the largest possible achievable PIR rate for an arbitrary linear code with the proposed protocol strongly depends
on the smallest possible value of $\frac{\kappa}{\nu}$ for which a PIR achievable rate matrix
$\mat{\Lambda}_{\kappa,\nu}$ exists. We also remark here that when we say that an MDS-PIR capacity-achieving matrix
$\mat{\Lambda}_{\kappa,\nu}$ exists, it does not necessarily require $(\nu,\kappa)=(n,k)$, but
$\frac{\kappa}{\nu}=\frac{k}{n}$.
A lower bound on the largest possible achievable PIR rate obtained from \cref{thm:PIRachievable-rate_code} and
\cite[Lem.~3]{KumarLinRosnesGraell17_1sub} is given as follows.
\begin{corollary}
  Consider a DSS that uses an $[n,k,d^\code{C}_\mathsf{min}]$ code $\code{C}$ to store $f$ files. Then, the PIR rate 
  \begin{equation*}
    \const{R}(\code{C})=\frac{\min\left(k,d^\code{C}_\mathsf{min}-1\right)}{n}
    \inv{\left[1-\Bigl(\frac{k}{k+\min\left(k,d^\code{C}_\mathsf{min}-1\right)}\Bigr)^f\right]}
  \end{equation*}
  is achievable. 
\end{corollary}

We remark that because every set of $k$ coordinates of an $[n,k]$ MDS code is an information set, we can construct $n$
information sets by cyclically shifting an arbitrary information set $n$ times, hence an MDS-PIR capacity-achieving
matrix $\mat{\Lambda}_{k,n}$ of an MDS code can be easily constructed. In other words, the proposed protocol with MDS
codes  is MDS-PIR capacity-achieving (see Corollary \ref{cor:MDSPIRcapacity-achieving-matrix}) and MDS codes are a class
of MDS-PIR capacity-achieving codes.

\section{MDS-PIR Capacity-Achieving Codes}
\label{sec:PIRcapacity-achiving-codes}

In this section, we provide a necessary and a sufficient condition for an arbitrary linear code to achieve the MDS-PIR capacity
$\const{C}_f$ (see \eqref{eq:PIRcapacity}) with the PIR protocol in \cref{sec:file-dep-PIRachievable-scheme}. 

\begin{theorem}
  \label{thm:general-d_PIRcapacity-achieving-codes}
  If an MDS-PIR capacity-achieving matrix exists for an $[n,k]$ code $\code{C}$, then
  \begin{equation}
    d_s^{\code{C}}\geq\frac{n}{k}s,\quad\forall\,s\in\Nat{k}.
    \label{eq:general-d_PIRcapacity-achieving-codes}    
  \end{equation}
\end{theorem}
\begin{IEEEproof}
  See the proof of \cite[Th.~3]{KumarLinRosnesGraell17_1sub}.
\end{IEEEproof}

Based on the necessary condition, it can be shown that the code $\code{C}$ in \cref{ex:n5k3_bad} is not MDS-PIR
capacity-achieving with the PIR protocol in \cref{sec:file-dep-PIRachievable-scheme}, since
$d_2^\code{C}=3<\frac{5}{3}\cdot 2$, i.e., it is impossible to find an MDS-PIR capacity-achieving matrix
$\mat{\Lambda}_{\kappa,\nu}$ for this code.


We have performed an exhaustive search for codes with parameters $k\in\Nat{n}$ and $n\in\Nat{11}$ (except for
$[n,k]=[10,5]$ and $[n,k]=[11,4\leq k\leq 7]$) and seen that for codes satisfying
\eqref{eq:general-d_PIRcapacity-achieving-codes}, there always exists an MDS-PIR capacity-achieving matrix. Therefore,
we conjecture that \eqref{eq:general-d_PIRcapacity-achieving-codes} is an if and only if condition for the existence of
an MDS-PIR capacity-achieving matrix.
\begin{conjecture}
  \label{conj:general-d_PIRcapacity-achiving-codes}
  An MDS-PIR capacity-achieving matrix $\mat{\Lambda}_{\kappa,\nu}(\code{C})$ with $\frac{\kappa}{\nu}=\frac{k}{n}$
  exists for an $[n,k]$ code $\code{C}$ if and only if \eqref{eq:general-d_PIRcapacity-achieving-codes} holds.
\end{conjecture}

In the following, we provide a sufficient condition for MDS-PIR capacity-achieving codes by using the code automorphisms
of an $[n,k]$ code\cite[Ch.~8]{macwilliamssloane77_1}.

\begin{theorem}
  \label{thm:sufficient_PIRcapacity-achieving-codes}
  Given an $[n,k]$ code $\code{C}$, if there exist $n$ distinct automorphisms $\pi_1,\ldots,\pi_n$ of $\code{C}$ such
  that for every code coordinate $j\in\Nat{n}$, $\{\pi_1(j),\ldots,\pi_n(j)\}=\Nat{n}$, then the code $\code{C}$ is an 
  MDS-PIR capacity-achieving code.
\end{theorem}
\begin{IEEEproof}
  See the proof of \cite[Th.~4]{KumarLinRosnesGraell17_1sub}.
\end{IEEEproof}

Using their known code automorphisms and \cref{thm:sufficient_PIRcapacity-achieving-codes}, it can be shown that the
families of cyclic codes and RM codes achieve the MDS-PIR capacity.
\begin{corollary}
  \label{cor:PIRcapacity-achiving_cyclic-RM-codes}
  Cyclic codes and RM codes are MDS-PIR capacity-achieving codes.
\end{corollary}

It can be easily shown that cyclic codes and RM codes satisfy the necessary condition of
\cref{thm:general-d_PIRcapacity-achieving-codes}.

\section{Conclusion}
\label{sec:conclusion}

We presented a PIR protocol for DSSs where data is stored using an arbitrary linear code for the case of noncolluding
nodes. By exploiting the information sets of the underlying storage code, an exact expression for the PIR rate of the
protocol was derived. Furthermore, a necessary and a sufficient condition for a code to be MDS-PIR capacity-achieving
were provided. We proved that cyclic codes and RM codes satisfy the sufficient condition and thus achieve the MDS-PIR
capacity with the proposed protocol.







\end{document}